# Deterministic fabrication of blue quantum emitters in hexagonal boron nitride

Angus Gale[1,#], Chi Li[1,#], Yongliang Chen[1], Kenji Watanabe[2], Takashi Taniguchi[3], Igor Aharonovich[1,4], Milos Toth[1,4*]

1. School of Mathematical and Physical Sciences, University of Technology Sydney, Ultimo, New South Wales 2007, Australia
2. Research Center for Functional Materials, National Institute for Materials Science, Tsukuba 305-0044, Japan
3. International Center for Materials Nanoarchitectonics, National Institute for Materials Science, Tsukuba 305-0044, Japan
4. ARC Centre of Excellence for Transformative Meta-Optical Systems, University of Technology Sydney, Ultimo, New South Wales 2007, Australia
# These authors contributed equally
*Email: Milos.Toth@uts.edu.au

*Hexagonal boron nitride (hBN) is gaining considerable attention as a solid-state host of quantum emitters from the ultraviolet to the near infrared spectral ranges. However, atomic structures of most of the emitters are speculative or unknown, and emitter fabrication methods typically suffer from poor reproducibility, spatial accuracy, or spectral specificity. Here, we present a robust, deterministic electron beam technique for site-specific fabrication of blue quantum emitters with a zero-phonon line at 436 nm (2.8 eV). We show that the emission intensity is proportional to electron dose and that the efficacy of the fabrication method correlates with a defect emission at 305 nm (4.1 eV). We attribute blue emitter generation to fragmentation of carbon clusters by electron impact and show that the robustness and universality of the emitter fabrication technique are enhanced by a pre-irradiation annealing treatment. Our results provide important insights into photophysical properties and structure of defects in hBN and a framework for deterministic fabrication of quantum emitters in hBN.*

Hexagonal boron nitride (hBN) is an attractive Van der Waals material due to its wide bandgap and chemical stability. It has been used widely as a thin dielectric layer in electronic devices and a protective cap for sensitive materials such as transition metal dichalcogenides. Recently, its quantum photonic properties have gained considerable attention as hBN has been found to host a variety of single photon emitters (SPEs) that span the ultraviolet (UV) to the near-infrared (IR) spectral range.[1-9]

hBN SPEs operate at room temperature and some have outstanding optical properties including high brightness, linear polarisation and access to the spin states of some of the defects.[10-12] The emitters are associated with intrinsic structural defects and extrinsic impurities in the lattice, and are found in many varieties of hBN samples, including those grown by the high pressure high temperature method, chemical vapor deposition and

MOVPE. [13-16] Numerous distinct defect-related quantum emitters have been identified in hBN. The various SPEs display a wide range of photophysical properties – including the emission wavelength, brightness, stability and spin properties. As a result, it is broadly accepted that numerous defect species are likely responsible for the emissions, but the atomic structures of most of these are speculative or unknown.

Consequently, numerous experiments have been conducted with the aim to engineer the emitters deterministically with reproducible emission properties. [4, 17, 18] These have met some success – for example, ensembles of the negatively charged boron vacancy ($V_B^-$) defect can be fabricated, but these emitters are dim and have so far not been isolated at a single defect level. [17-19] Conversely, emitters in the visible spectral range can be engineered at a single defect level, but fabrication methods that are deterministic both spatially and spectrally remain elusive. This limits not only studies of defect structure but also efforts at scalable integration of hBN emitters in photonic circuits and devices. As a result, there is a need for improved emitter fabrication methods and detailed fundamental studies of specific fluorescent defects that can be engineered on demand in a wide range of hBN samples.

Here, we present a robust electron beam technique for site-specific engineering of defects emitting at 436 nm (2.8 eV), referred to as blue emitters throughout the manuscript. For the first time, we show highly controlled, dose-dependent engineering of emitters and correlate the efficacy of the method with a spectral signature of hBN flakes. Employing nanoscale CL spectroscopy, we show that the effectiveness of the blue emitter generation process correlates with a well-studied carbon-related UV center at 305 nm (4.1 eV).[3, 13, 20-26] We attribute blue emitter formation to decomposition of carbon clusters in hBN by electron impact, and show that the efficacy of the emitter fabrication method can be improved by an annealing treatment. Our findings provide a new approach for deterministic fabrication of quantum emitters in hBN with nanoscale resolution.

To engineer emitters using electron beams, two sources of hBN were studied: flakes were exfoliated from hBN grown using the high-pressure and high-temperature (HPHT) method ("Type I"), and from hBN that was purchased from a commercial supplier ("Type II"). The flakes were exfoliated onto $SiO_2$/Si substrates and annealed in $N_2$ at 1000 °C. The function of this pre-irradiation annealing treatment is discussed below. Importantly, no additional annealing was done after electron irradiation. Optical characterisation was performed using a cathodoluminescence (CL) system installed on the scanning electron microscope (SEM) used to fabricate emitters, and an *ex-situ* custom-built PL setup. CL analysis was performed both during electron beam processing, as well as before/after processing using a low current density electron beam that modifies samples minimally during analysis. See the Methods section for additional details.

A schematic of the CL setup is shown in Figure 1a. CL is excited by high energy (keV) electrons that can excite both defects as well as inter-band transitions and is therefore complementary to PL performed using a sub-bandgap excitation source. A parabolic mirror reflects and collimates CL into a setup that enables spectroscopic and correlation measurements.

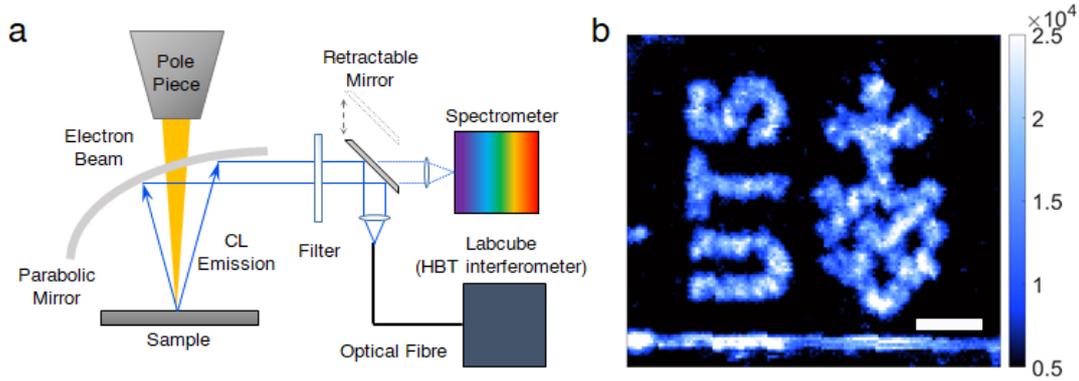

*Figure 1. Direct-write electron beam fabrication of blue emitters in hBN. (a) Schematic illustration of the electron beam setup used to process hBN and perform CL analysis. (b) Confocal PL map of the UTS emblem patterned by electron beam irradiation of hBN. The map was collected using a 60 nm bandpass filter centred on 460 nm. The scale bar represents 5 μm.*

We start by demonstrating the spatial control and consistency of the emitter fabrication method by writing the University of Technology Sydney emblem into a Type II hBN flake using a 5 keV electron beam (0.8 nA). Figure 1b shows a confocal PL map of the emblem acquired using the blue emission generated using the electron beam. The emission was collected using a 60 nm bandpass filter centered on 460 nm. The fabrication resolution is limited by the diameter of the electron beam, and electron scattering in the sample.

Normalized, room temperature CL and PL spectra of the blue emission generated by the electron beam are shown in Figure 2a. The spectra consist of a dominant peak at 436 nm and two satellite peaks at 462 nm and 491 nm. We attribute these to a ZPL and a PSB comprised of two phonon replicas, consistent with prior reports. [4, 27-29] Differences between the CL and PL spectra are caused by a lower resolution of the CL spectrometer (see SI, Figure S1). A low temperature PL spectrum is shown in Figure 2b. It shows expected reductions in peak widths and in the relative intensity of the PSB, and the spectrum is consistent with prior studies of emitters in hBN [30, 31] – the apparent asymmetry in the ZPL is attributed to a ZA (out-of-plane acoustic) phonon replica detuned from the ZPL by 10 meV, and the longer wavelength emissions are consistent with the phonon modes of hBN. [28, 32]

Photon emission statistics were analysed using both CL and PL. The blue curve in Figure 2c is a second-order CL autocorrelation function, $g^{(2)}(\tau)$, measured using a 5 keV (8 pA) electron beam, from an ensemble of the blue emitters. It shows significant bunching, with a $g^{(2)}(0)$ value of 4.53 ± 0.03. Photon bunching is typical of CL emissions from quantum emitter ensembles excited using a low current electron beam, and has previously been ascribed to simultaneous excitation of an ensemble by plasmon decay. [33] The extent of bunching decreases with increasing electron beam current, as is illustrated in the SI, Figure S2. A mean emitter lifetime of 2.11 ± 0.03 ns was extracted from a single exponential fit of the CL $g^{(2)}$ function. This value is consistent with and typical of the excited state lifetimes of emitters in hBN. [34] A second-order PL autocorrelation measurement from a single blue emitter generated by an electron beam is shown in red in Figure 2c. It shows antibunching with a $g^{(2)}(0)$ value of 0.50 ± 0.14, which decreases to below 0.1 after background correction (see SI,

Figure S3). A PL decay measurement that was obtained at 70 K and fitted with a single exponential function, yielding an excited state lifetime of 1.85 ns, consistent with the mean value of ~2.1 ns obtained from the CL autocorrelation function of an ensemble of these emitters. The optical measurements in Figure 2 summarise the basic properties of the blue emitters generated by electron irradiation of hBN, and illustrate the non-classical nature of photon emission from ensembles and individual emitters excited by electrons and photons, respectively. Characterisation of quantum emitters by both CL and PL is important for future applications of hBN quantum emitters in nanophotonics and nanoplasmonics, where CL and electron beam techniques play an increasingly important role. [35]

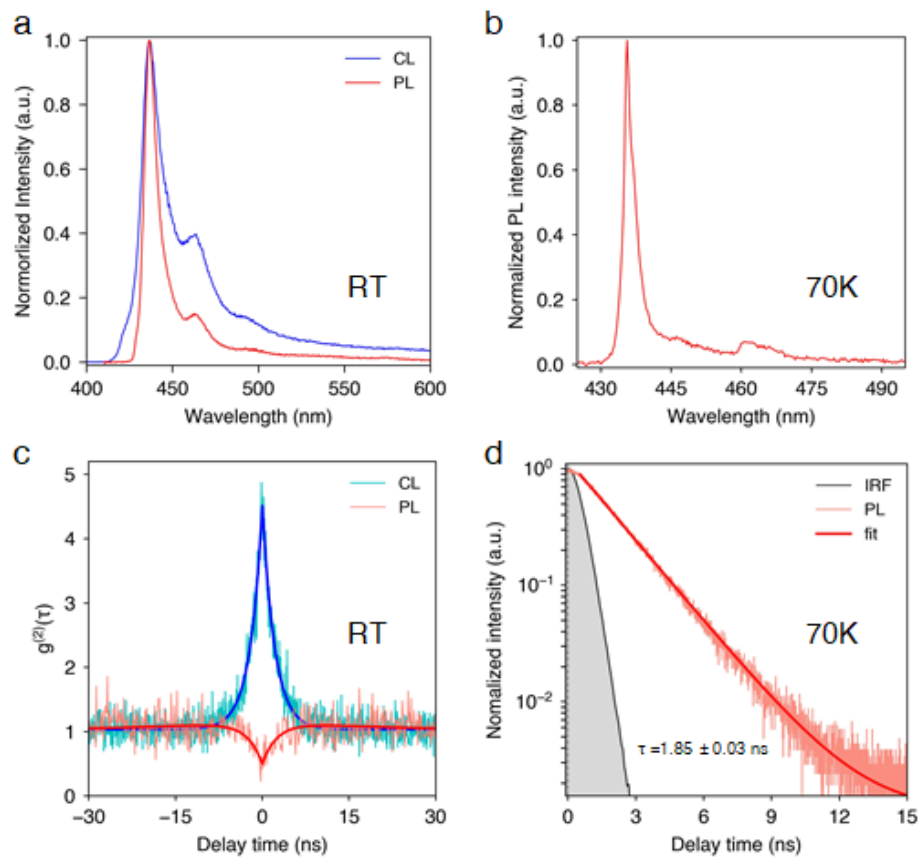

*Figure 2. Optical properties of blue emitters fabricated by electron irradiation of hBN. (a) CL (blue) and PL (red) spectra collected at room temperature. (b) Cryogenic PL spectrum collected at 70 K. (c) Second-order CL autocorrelation function (blue) of an ensemble, and a PL autocorrelation function (red) from a single emitter, measured at room temperature. The intensity at zero delay time is 4.53±0.03 and 0.50±0.14, respectively. Background correction was not employed in either case. A lifetime of 2.11±0.03 ns was deduced from a single exponential fit of the CL $g^{(2)}$ curve. (d) PL decay curve obtained using a 405 nm pulsed laser, yielding an excited state lifetime of 1.85±0.03 ns. The excitation pulse width and repetition rate were 45 ps and 20MHz, respectively. A 430 nm longpass filter was used for the PL measurement in (a), and a 60 nm bandpass filter centered on 460 nm was used in (b-d).*

We now turn to dynamics of the blue emitter fabrication process. Figure 3a and b shows CL and PL maps of a 3x3 spot array fabricated in an annealed Typed II hBN flake as a function of electron beam exposure time. The electron dose was varied from 6.2x10$^9$ to 1.6x10$^{12}$, as is shown in the legend of Figure 3c. The CL map was generated by integrating the CL intensity between 430 and 480 nm, and the PL map was acquired from the same region using a 460±30nm bandpass filter. Both maps show enhanced fluorescence intensity at each spot irradiated by the electron beam. Figure 3c shows PL spectra recorded from the nine spots, and the integrated PL intensity is plotted as a function of electron dose in Figure 3d. The PL intensity scales linearly with electron dose, over the range used in the experiment. Notably, such controlled, dose-dependent generation has not been demonstrated previously for quantum emitters in hBN.

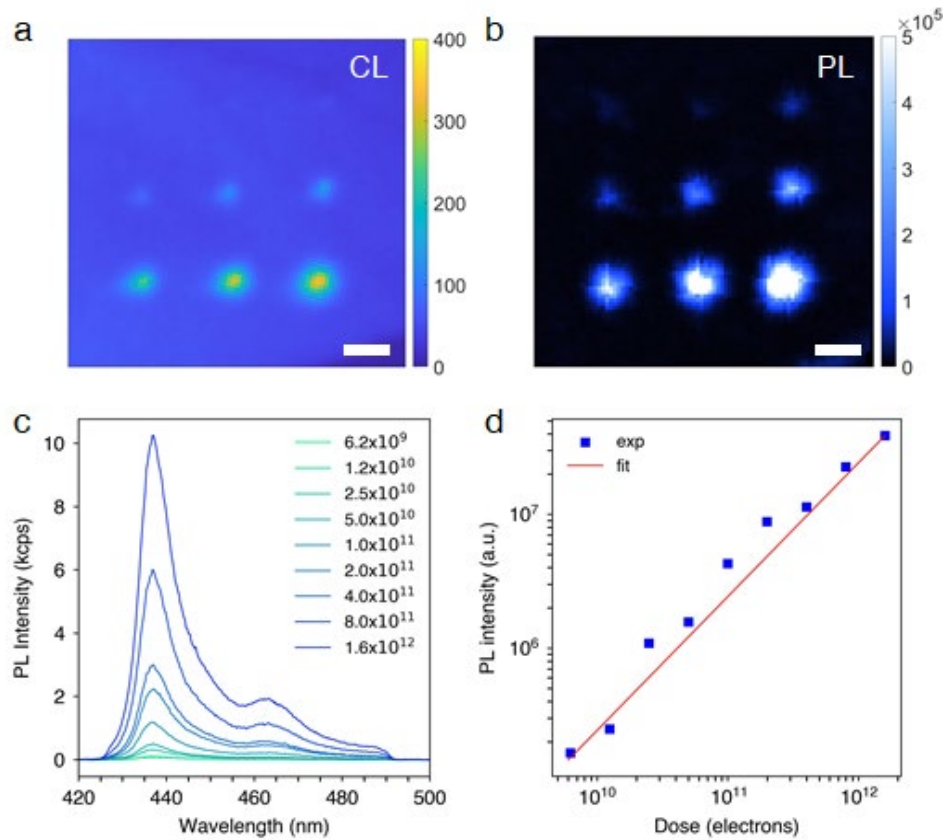

*Figure 3. Dynamics of the emitter fabrication process – emission intensity versus electron dose. (a,b) CL and PL maps of a 3x3 emitter array generated using a 1 nA stationary electron beam as a function of exposure time. From top left to bottom right, the irradiation times are 1, 2, 4, 8, 16, 32, 64, 128 and 256 s. The CL map was extracted from a hyperspectral map by integrating the CL intensity in the spectral range 430 - 480 nm. The scale bars represent 2 μm. (c) PL spectra obtained from the nine spots seen in the maps. The legend specifies the electron dose used to generate each spot. (d) PL intensity versus electron dose obtained by integrating the spectra in (c).*

We note that the analysis in Figure 3d was performed using PL spectra produced using a sub-bandgap excitation source, and that the CL signal is not appropriate for this analysis. In CL, carriers are excited across the bandgap, recombination of electron-hole pairs is a competitive process and CL intensity therefore does not, in general, scale linearly with defect concentration.[36] We emphasize that the data in Figure 3 were acquired from hBN that was annealed at 1000 °C before electron beam processing, and that no additional annealing (or any other processing) was performed after electron irradiation. The linear dependence on electron dose seen in Figure 3d suggests that the blue emitter generation rate is not rate-limited by diffusive mass transport through the hBN lattice, as is discussed below. The dose dependence and general reproducibility of the method within a flake also suggest that the defects are abundant and distributed relatively uniformly throughout the hBN lattice.

To elucidate the blue emitter generation process further, we study the processes using hBN samples that were grown by two methods, and optionally annealed before electron irradiation. The function of the pre-annealing treatment is demonstrated in Figure 4 by CL spectra from Type I, and Type II hBN that were irradiated in either their as-grown state (black lines) or after annealing (red curves). The high energy peak at 216 nm is the near-bandgap emission of hBN,[37] and the starred peak at 432 nm is the corresponding second order peak. The next prominent feature in the spectra is a strong, narrow mid-near UV emission at 305 nm (4.1 eV) – the ZPL of a well-documented defect[3, 13, 24, 25] that has been ascribed to carbon dimers in the literature,[20-23] and associated TO phonon replicas at 320 nm (3.9 eV) and 334 nm (3.7 eV). A third, less intense phonon replica is seen as a shoulder at 351 nm (3.5 eV). The last peak of interest in the spectra is the blue emission generated by electron irradiation with a ZPL at 436 nm (2.8 eV), [4, 27] which is enlarged in the panels on the right-hand side of Figure 4.

Flakes exfoliated from Type I hBN (grown by the HPHT method) were found to exhibit a strong near band edge emission, and to fall into two general categories, designated Type Ia and Type Ib in Figure 4a,b. Type Ia flakes do not show the 305 nm UV emission (and the associated phonon sideband), but instead show broad emissions centered on 259, 299 and 350 nm. Type Ib, flakes are characterised by the 305 nm UV emission (and the associated phonon sideband). Such differences between fluorescence spectra of Type I hBN have been observed previously and attributed tentatively to intrinsic defects,[38] and variations in carbon and oxygen content. [13]

We observe a strong correlation, illustrated by the CL spectra in Figure 4, between the 305 nm UV emission (~ 4.1 eV) and the efficacy of electron beam irradiation at generating the 436 nm blue emitters. Specifically, the irradiation technique is ineffective in flakes exfoliated from Type Ia hBN which do not show the 305 nm UV emission, and highly effective in Type Ib flakes which do show the UV emission. In Type Ia flakes, the electron beam changes the intensities of broad emissions centered on ~259, 299 and 350 nm, but it is very ineffective at generating the 436 nm blue emission (Figure 4a). Conversely, in Type Ib flakes, the electron beam does consistently generate the 436 nm emission, as is illustrated by Figure 4b. This correlation provides a strong indication that the 436 nm blue emitters are also associated with carbon, consistent with recent modelling studies. [20, 23]

The variability between Type Ia and Ib samples (which were exfoliated from a single crystal of hBN), can be suppressed by the 1000 °C annealing treatment performed in $N_2$ prior to electron exposure. The vast majority of annealed Type I flakes are characterised by CL spectra dominated by the 305 nm UV emission, and we found electron irradiation to generate the

436 nm emission consistently and reliably in these flakes, as is illustrated in Figure 4c. We note that the 216 nm near-bandgap CL emission is very weak in annealed Type I flakes.

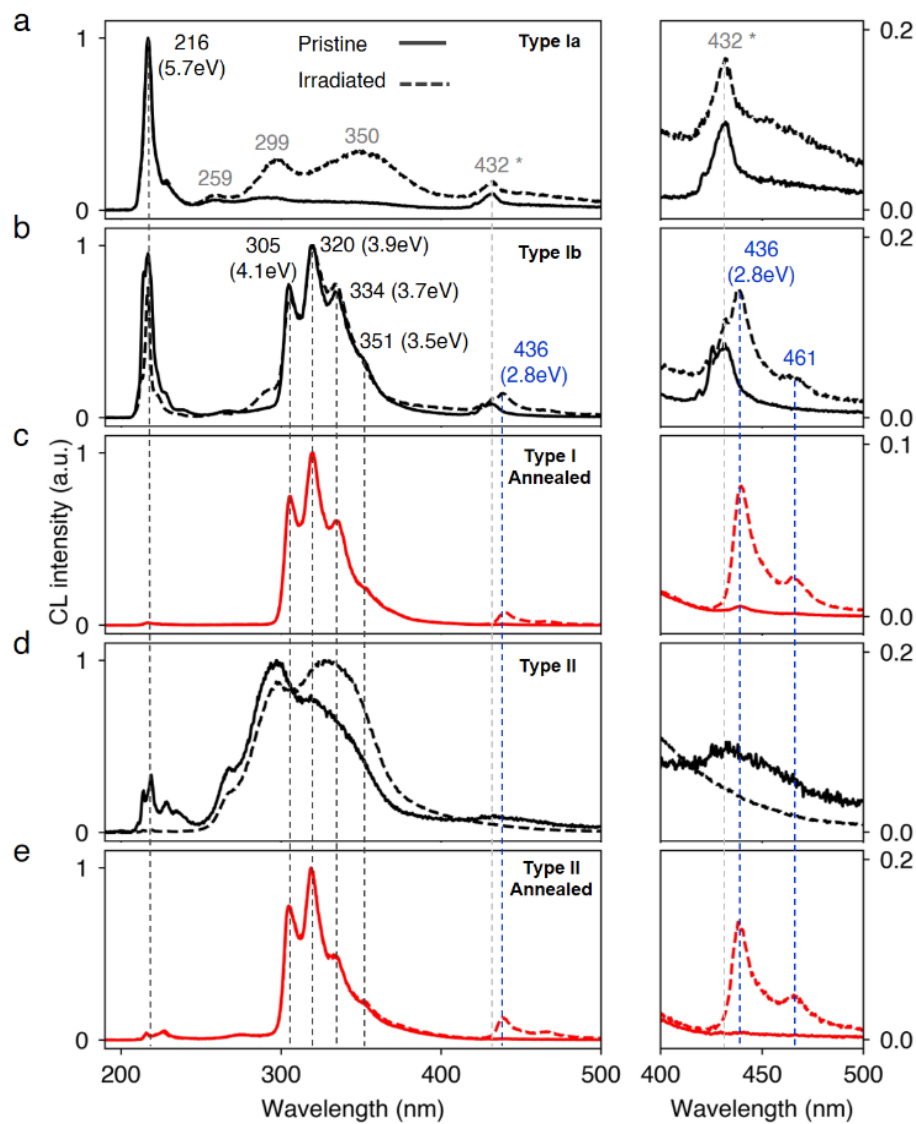

*Figure 4. CL characterisation of hBN before and after processing by an electron beam.* (a - e) *Normalised CL spectra of five hBN samples before (solid lines) and after (dashed lines) electron beam processing. Black and red curves indicate samples that were as-grown and annealed at 1000 °C, respectively. Peak positions are indicated on the plots in nanometres and electronvolts (eV). The asterisk indicates a second order peak at 432 nm. The right panel is a close-up of the spectral range containing the 436 nm blue emission (and a phonon replica at 461 nm) generated by electron irradiation of hBN samples that show a sharp UV emission at 305 nm (and associated phonon replicas at 320, 334 and 351 nm).*

Similar behaviour was observed in commercially-sourced Type II hBN. Typical CL spectra obtained from flakes of this material before and after electron irradiation are shown in Figure 4d. In as-grown samples, the spectra do not show the sharp 305 nm UV emission, and the electron irradiation treatment is consistently ineffective at generating the 436 nm blue emission. However, as is illustrated by Figure 4e, the 1000 °C annealing treatment is again highly effective at producing the 305 nm emission, and it greatly enhances the efficacy of electron irradiation at generating the 436 nm blue emitters.

We now discuss potential mechanisms behind the annealing and electron irradiation treatments. In general, annealing can create new defects, cause the diffusion and restructuring of existing defects, introduce impurities/dopants, and shift the Fermi level which can alter the charge states of existing defects. We note that the annealing treatment was ineffective at generating the 305 nm UV emission in a small subset of Type I hBN flakes (and electron beam irradiation was ineffective at generating the 436 nm blue emitters in this small subset of flakes both before and after annealing). These flakes are characterised by CL spectra dominated by an intense near-bandgap emission both before and after annealing, which is an indicator of highly pristine, high quality hBN. [37-39] We therefore conclude that the primary function of the annealing treatment is not to introduce new defects/impurities, but rather to restructure existing defects, which likely involve impurity atoms that were distributed unevenly in the as-grown hBN crystal used for exfoliation.

Returning to flakes that were annealed "successfully", yielding CL spectra dominated by the 305 nm UV emission (Figure 4 c and e), we note that the annealing increased not only the flake sensitivity to the electron beam emitter fabrication process, but also uniformity within the flakes – meaning that annealing increased the consistency of a linear relationship between electron dose and intensity of the 436 nm blue emission (illustrated by Figure 3d). This implies that annealing homogenised the concentrations of defects responsible for the 305 nm (4.1 eV) UV emission. In the literature, this UV emission has been associated with carbon impurities in hBN, [13, 26] and recent modelling studies suggest the $C_BC_N$ carbon dimer as the underlying atomic structure. [20-23] Hence, based on this picture, our results indicate that the annealing treatment gives rise to the formation of $C_BC_N$ dimers in carbon-containing hBN flakes.

Next, we turn to electron irradiation. Energetic electrons can generate new defects and restructure existing defects. At the electron energies employed here, the primary mechanism is not the displacement of nuclei via momentum transfer (i.e., knock-on)[40, 41] but instead bond-breaking caused by electron-electron collisions, and transport assisted by drift of ionised atoms in the presence of electric fields generated in a dielectric by an electron beam.[42, 43] Bearing this in mind, we focus on our observation that electron irradiation is effective at generating the 436 nm blue emitters only in hBN flakes with CL spectra characterised by the 305 nm UV emission. Specifically, we exclude the possibility that the blue emitters are generated purely because new defects are created by the electron beam. Whilst defect creation can occur, a clear prerequisite for blue emitter generation is presence of the UV defects, and it is reasonable to suggest that these defects are modified by the electron beam. We note also that the PL intensity of the 436 nm emission scales linearly with electron dose, over at least two orders of magnitude ($10^{10}$ to $10^{12}$ electrons, see Figure 3d), indicating that atomic drift/diffusion through the hBN lattice is not the rate-limiting process in annealed hBN flakes. Instead, we consider two possibilities – electron irradiation may modify the defect charge state or atomic structure. The former can proceed directly by electron-induced charge transitions in hBN, or indirectly *via* band bending[44] caused by electron beam modification of the surface. The latter is the more likely in the light of modelling studies which indicate that the $C_BC_N$ dimer is the source of the 305 nm UV emission, and that calculated PL spectra of isolated and closely-spaced substitutional carbon atoms resemble the 436 nm blue emission.[20, 23] Hence, the blue emitters likely form *via* restructuring of the carbon dimers through a bond breaking-restructuring process initiated by electron impact.

We have developed a robust approach for site-specific engineering of blue quantum emitters in hBN that involves a high temperature anneal, pre-characterisation by CL spectroscopy and an electron irradiation treatment. The fabricated defects emit at 436 nm (2.8 eV) and the ability to generate these blue emitters by an electron beam correlates with the pre-existence of carbon-related defects that emit at 305 nm (4.1 eV). Our results improve present understanding of the photophysical properties of hBN, and on the effects of electron irradiation on defects in hBN. The blue emitter fabrication method is appealing for both fundamental studies of quantum emitters in hBN and for their integration in photonic nanostructures – it has high spatial resolution, generates a specific defect species characterised by an emission wavelength of 436 nm, and the emission intensity scales linearly with electron dose.

**Methods**
**Sample Preparation**
$Si/SiO_2$ (285 nm oxide) substrates were first solvent-cleaned and ultrasonicated in acetone then isopropyl alcohol before being dried in nitrogen. The substrates were further cleaned for 30 mins in a commercial ozone cleaner before exfoliating hBN using PDMS. Two different hBN bulk crystals were used: HPHT hBN from the National Institute for Materials Science (NIMS) and commercially available hBN from HQ Graphene. To remove unwanted residue after exfoliation the samples were annealed at 500 °C in air for 12 hours before being ozone cleaned for 30 minutes. A tube furnace was used to anneal samples at 1000 °C for 1 hour in a nitrogen atmosphere (1 Torr, 50 SCCM).

**Emitter Creation**
Blue emitters were created using an FEI DB235 Dual Beam FIB/SEM microscope. Beam energies of 5-10 keV and currents ranging from 8.0 pA - 4.5 nA were used for all experiments. The spectra and time resolved data in Figure 2 were collected from irradiated spots on annealed HQ Graphene hBN. CL spectra were monitored during emitter creation and used to determine the irradiation time. The arrays in Figure 3 were patterned on annealed NIMS hBN using a 10 keV, 1.0 nA beam on squares of area 20 x 20 nm. The dose was controlled by modulating the total irradiation time from 1 - 256 seconds. Irradiations in Figure 4 were performed using a 10 keV, 1.0 nA stationary beam (240 seconds for a total dose of 1.50 x $10^{12}$ electrons). The UTS Logo in Figure 1b was patterned on annealed HQ Graphene hBN using a Thermo Fisher Scientific Helios G4 Dual Beam microscope using a beam energy of 5.0 keV and a beam current of 0.8 nA. The area was patterned for a total of 20 minutes.

**Cathodoluminescence measurements**

CL mapping, spectra and time correlated measurements were collected using a Delmic SPARC system with a 13 mm parabolic mirror. Data was collected with a 10 keV beam energy and 0.28 - 1.0 nA beam current. CL emission was sent to a spectrometer (Andor Kymera 193i) using a slit width of 150 μm. The spectrometer was equipped with a 300 lines/mm grating for CL mapping and individual spectra. A pixel size of 200 nm and dwell time of 10 ms were used for the map in Figure 3a. For single spectra of the SPEs in Figure 2a, a 430 nm long-pass filter was placed in the optical path to remove second order features from high energy CL emission. For time resolved measurements in Figure 2c the CL emission was directed to the Delmic LAB Cube. This encompases a Hanbury Brown and Twiss setup consisting of a 50/50 beam splitter and two photomultiplier tube detectors (Hamamatsu R943-02). The emission was filtered using a 460±30 nm band-pass filter. The electron beam was scanned across an area of 150 x 130 nm containing an existing SPE ensemble for 1 hour using a beam energy of 5.0 keV and beam current of 8.0 pA. Spectra presented in Figure 4 were collected at the beginning and end of the irradiations outlined above. Integration times of 1 second for non-annealed hBN and 10 ms and 100 ms were used for annealed NIMS and HQ Graphene samples respectively. Delmic Odemis software was used for all data collection.

**Photoluminescence measurements**

The room temperature PL measurements in Figures 1-3, were conducted using a lab-built confocal microscope. Briefly, a 405-nm continuous-wave (CW) laser (PiL040X, A.L.S. GmbH) excites the sample via a 100× objective (NA ~0.9, Nikon). The reflection/fluorescence was filtered with a dichroic mirror (long-pass 405 nm) and collected either by avalanche photodiode single photon detectors (APDs, Excelitas Technologies) or a spectrometer (Princeton Instruments, Inc.).

The cryogenic PL measurements in Figure 2 were performed with a lab-built confocal setup equipped with an open-loop cryostat (a ST500 cryostat, Janis) containing flowing liquid Nitrogen. A three-dimensional piezostage (ANPx series, attocube Inc.) is located inside the cryostat to adjust the position of the samples. A cryogenic temperature controller (335, Lakeshore) was used to adjust the temperature of the samples. A thin quartz window enables optical access to the samples. A 405-nm CW laser was used to excite the samples through a 100× objective (NA, 0.9; TU Plan Fluor, Nikon). Back-collected fluorescence was filtered and directed either to an APD or a spectrometer (SR303I, Andor).

For time-resolved PL spectroscopy in Figure 2, a pulsed 405-nm laser (pulse width 45 ps) with 20 MHz repetition rate was used as the excitation source. A correlator (PicoHarp300, PicoQuant) was used to synchronize PL emission.

**Acknowledgements**

The authors acknowledge financial support from the Australian Research Council (CE200100010, DP190101058) for financial support. The authors thank Mark Lockrey for his assistance with the CL system and John Scott for fruitful discussions. Kenji Watanabe and Takashi Taniguchi acknowledge support from the Elemental Strategy Initiative conducted by the MEXT, Japan, Grant No. JPMXP0112101001, JSPS KAKENHI Grant No. JP20H00354 and the CREST (JPMJCR15F3), JST.